\documentclass[12pt,fleqn]{article}
\usepackage{amsfonts,amsmath,amsthm, enumerate}
\newcommand{\lm}{\lambda}

\newcommand{\R}{\mathbb{R}}

\newcommand{\N}{\mathbb{N}}
\newcommand{\Z}{\mathbb{Z}}
\newcommand{\C}{\mathbb{C}}
\newcommand{\E}{\mathbb{E}}
\newcommand{\PP}{\mathbb{P}}
\newcommand{\T}{\mathbb{T}}
\newcommand{\gm}{\gamma}
\newcommand{\al}{\alpha}
\newcommand{\bra}{\langle}
\newcommand{\ket}{\rangle}
\newcommand{\be}{\begin{equation}}
\newcommand{\ee}{\end{equation}}
\newcommand{\bea}{\begin{align}}
\newcommand{\eea}{\end{align}}

\newcommand{\og}{\omega}

\newtheorem{Lemma}{Lemma}[section]
\newtheorem{Theorem}{Theorem}[section]
\newtheorem{Corollary}{Corollary}[section]
\newtheorem{Proposition}{Proposition}[section]

\begin{document}
\title{Lyapunov Exponents for Unitary Anderson Models}
\author{Eman Hamza and G\"unter Stolz\footnote{partially supported through US-NSF grant DMS-0245210}\\
University of Alabama at Birmingham\\
Department of Mathematics
 CH 452\\
Birmingham, AL 35294-1170\\ U.S.A. }
\date{}
\maketitle \abstract{We study a unitary version of the
one-dimensional Anderson model, given by a five diagonal
deterministic unitary operator multiplicatively perturbed by a
random phase matrix. We fully characterize positivity and
vanishing of the Lyapunov exponent for this model throughout the
spectrum and for arbitrary distributions of the random phases.
This includes Bernoulli distributions, where in certain cases a
finite number of critical spectral values, with vanishing Lyapunov
exponent, exists. We establish similar results for a unitary
version of the random dimer model.}

\setcounter{equation}{0}
\section{Introduction}
Unitary operators arise naturally in quantum mechanics as the time
evolution of the Hamiltonian in solving the time dependent
Schr\"{o}dinger equation. In particular, for Hamiltonians with
periodically time-dependent potentials, the spectral properties of
the monodromy operator (the unitary operator giving the evolution
over one time period) are the central object of mathematical
investigations (e.g.~\cite{b},~\cite{c},~\cite{h}). A model using
random unitary operators was used in~\cite{bb} to study the single
particle behavior of an electron in a small one-dimensional metal
ring in the presence of a large uniform electric field generated
by a linearly ramped magnetic flux. The unitary operator used in
describing this system is characterized by a five diagonal band
structure. Motivated by this model, the spectral analysis of a
class of random unitary operators with similar band structure was
undertaken in~\cite{bhj},~\cite{j1},~\cite{j2} and~\cite{hjs}.

This class of operators on $l^2(\Z)$ can be written (up to a
unitary equivalence) in the form
\begin{equation} \label{mainmodel}
U_\omega=D_\omega S,
\end{equation}
where $S$ is a unitary operator with a five-diagonal matrix
representation. $S$ depends on a parameter $t\in (0,1)$ which
controls the size of its off-diagonal elements and takes the role
of a disorder parameter for $U_{\omega}$, see Section~\ref{model}
below. $D_\omega$ is a diagonal matrix of random phases,
diag$\{e^{-i\theta_k^\omega}\}$. For our application,
$\{\theta_k^\omega:k\in\Z\}$ is a sequence of i.i.d.\ random
variables on the one-dimensional torus $\mathbb{T} =\R/2\pi\Z$
with a non-trivial probability distribution $\mu$, i.e.
supp$\,\mu$ has two or more elements. $U_\omega$ can be considered
as a ``unitary Anderson-type model", where $S$ plays the role of
the free Laplacian and where the perturbation is introduced via
multiplication rather than addition to ensure that the resulting
operator is still unitary. Indeed, due to the band structure of
the operator, the generalized eigenvectors can be studied using
complex $2\times 2$ transfer matrices. This formalism allows to
introduce the Lyapunov exponent $\gamma(\lambda)$, where
$e^{i\lambda}$ is the spectral parameter, see equation
(\ref{def.lyp.exp.}) below. Due to the fact that the transfer
matrices have determinants of unit modulus, the Lyapunov exponent
is almost surely non-negative. This also allowed to prove a
unitary version of the Ishii-Pastur Theorem, i.e.\
Theorem~\ref{IP} below, and to deduce the absence of absolutely
continuous spectrum in case of uniform distribution
$\mu$~\cite{bhj}. In~\cite{hjs} it was shown that the spectrum of
$U_\omega$ is almost surely pure point with exponentially decaying
eigenfunctions if the distribution $\mu$ has a non-trivial
absolutely continuous component and non-empty interior. This holds
for arbitrary value of the disorder parameter $t$. Thus, up to
this point, the results for one-dimensional unitary Anderson
models are in close analogy to those known for the self-adjoint
Anderson model.

Further results on unitary Anderson models were found in
\cite{j1}, where the density of states is studied, and in
\cite{j2}, which develops a fractional moment approach to prove
localization for multi-dimensional analogs of unitary Anderson
models.

The main goal of this paper is to further investigate the model
(\ref{mainmodel}) and to fully determine the set $\{\lambda \in
\mathbb{T}: \gamma(\lambda)>0\}$ for arbitrary disorder $t$ and
arbitrary distribution $\mu$, including singular and, in
particular, Bernoulli distributions.

For most choices of $\mu$ we find that $\gamma(\lambda)>0$ for all
``quasi-energies'' $\lambda \in \mathbb{T}$, see
Theorem~\ref{pos.lyp}. However, there is one exceptional
situation: If $\mu$ is a Bernoulli measure supported on two
diametrically opposed points, i.e.\ supp$\,\mu = \{a,b\}$,
$|a-b|=\pi$, then there exist two critical quasi-energies $\lambda
=-a$ and $\lambda=-b$ at which the Lyapunov exponent vanishes,
while it is positive for all other values of $\lambda$
(Theorem~\ref{picase}). In fact, we show in the proof of
Theorem~\ref{picase} that at the anomalies $\lambda=-a$ and
$\lambda=-b$ the transfer matrices $T_n(\omega,\lambda)$ satisfy
the asymptotics
\begin{equation} \label{transferasymp}
\frac{1}{n} \E\left((\ln \|T_n(\omega,\lambda)\|)^2\right)
\longrightarrow C>0,
\end{equation}
i.e., roughly, $\|T_n(\omega,\lambda)\| \sim e^{(Cn)^{1/2}}$.

As there are no more than two critical values of $\lambda$, the
unitary version of the Ishii-Pastur Theorem shows that the unitary
Anderson model (\ref{mainmodel}) almost surely has no absolutely
continuous spectrum, irrespective of the underlying probability
measure and disorder. We expect that methods such as those used in
\cite{CKM} can be adjusted to also show that $U_{\omega}$ almost
surely has pure point spectrum with exponentially decaying
eigenfunctions, i.e.\ is spectrally localized, but we haven't
carried out the details of this.

It is interesting that the structure of Lyapunov exponents for the
unitary Anderson model is richer than for the self-adjoint
one-dimensional Anderson model. For the latter it has been long
known that the Lyapunov exponent is positive at all energies for
all non-trivial single site distributions of the random potential,
e.g.\ \cite{cl}.

In a more general class of self-adjoint Anderson-type models it
has been shown that the existence of critical energies with
vanishing Lyapunov exponents can lead to the co-existence of
spectral localization and suitable forms of dynamical
delocalization, e.g.\ \cite{jss}. The simplest self-adjoint model
which shows this phenomenon is the so-called dimer model
\cite{dlw, dg}, in which the random phases appear in the form of
identical neighboring pairs. The typical anomalies encountered in
the dimer model are stronger than those in Theorem~\ref{picase}
below in the sense that transfer matrices, rather than satisfying
(\ref{transferasymp}), are uniformly bounded in $n$. In
$\cite{jss}$ it is shown that this leads to super-diffusive
transport, while, to our knowledge, the dynamical effects caused
by an anomaly as in (\ref{transferasymp}) have not been studied
(they should be much weaker, if detectable at all).

In Section~\ref{dimer} we study a unitary version of the dimer
model, where we can show that the Lyapunov exponent is positive
away from at most finitely many critical values. However, for the
dimer model with Bernoulli distributed phases, i.e.\ supp$\,\mu =
\{a,b\}$, and such that $|a-b|$ is in the spectrum of $S$, there
are two critical values where transfer matrices are of the type
studied in \cite{jss}, in particular they are bounded in $n$.

While we do not carry out a complete study of spectral and
dynamical localization properties of unitary Anderson models, in
Section~\ref{contlyap} we state one more result which has been
relevant in this context in the self-adjoint case, namely
continuity of the Lyapunov exponent in $\lambda$ away from the
critical quasi-energies. This is proven by a rather direct
adaptation of the proof in the self-adjoint case, e.g.\ \cite{cl}.

Let us finally mention that unitary operators with the same band
structure as $S$ and $U_{\omega}$ above also arise in the form of
so-called CMV-matrices in the study of orthogonal polynomials on
the unit circle, e.g.\ \cite{s,cmv}. In this setting, a definition
of Lyapunov exponents different from ours arises naturally
(applicable for example for the case of i.i.d.\ random Verblunsky
coefficients), see Section~10.5 in \cite{s}. However, it can be
shown that both definitions lead to the same value \cite{lenz} (if
either one of the two Lyapunov exponents exists).

%%%%%%%%%%%%%%%%%%%%%%%%%%%%%%%%%%%%%%%%%%%%%%%%%%%%%%
\section{The Model}\label{model}
Analogous to the self-adjoint case, we look at a random unitary
operator as a random perturbation of a deterministic (``free'')
unitary operator. The model and the results presented in this
section can be found in~\cite{bhj},~\cite{hjs},~\cite{j1}
and~\cite{j2}, see there for details and additional motivation and
results. Motivated by~\cite{bb}, we choose the free unitary
operator $S$ on $l^2(\Z)$ with band structure \be\label{s0}
S={\begin{pmatrix}\ddots & rt & -t^2& & & \cr
              & r^2& -rt  & & & \cr
              & rt & r^2 & rt & -t^2& \cr
              & -t^2 &-tr & r^2& -rt& \cr
              & & & rt &r^2 & \cr
              & & & -t^2& -tr&\ddots \end{pmatrix}},
\ee where the position of the origin in $\Z$ is fixed by $\bra
e_{2k-2},S e_{2k}\ket =-t^2$, with $e_k$ ($k\in\Z$) denoting the
canonical basis vectors in $l^2(\Z)$. The real parameters $t$ and
$r$ are linked by $r^2+t^2=1$ to ensure unitarity. Due to unitary
equivalence it suffices to consider $0\leq t,r \leq 1$. Thus $S$
is determined by $t$. We shall sometimes write $S(t)$ to emphasize
this dependence. Excluding trivial special cases, we assume
$0<t<1$. The spectrum of $S(t)$ is purely absolutely continuous
and is given by the arc $$ \sigma(S(t))=\Sigma(t)=
\{e^{i\vartheta}: \vartheta \in [-\arccos(1-2t^2),
\arccos(1-2t^2)]\}, $$ which is symmetric about the real axis and
grows from the single point $\{1\}$ for $t=0$ to the entire unit
circle for $t=1$.

The random perturbation is then introduced via multiplication by a
diagonal matrix
\be\label{d}D_\omega=\text{diag}\{e^{-i\theta_k^\omega}\},\ee with
$\{\theta_k^\omega:k\in\Z\}$ a sequence of i.i.d.\ random
variables on the torus $\mathbb{T} = \R/2\pi\Z$. More precisely,
we introduce the probability space
$(\Omega,\mathcal{F},\mathbb{P})$, where $\Omega$ is identified
with ${\mathbb{T}^\Z}$, $\mathcal{F}$ is the $\sigma$-algebra
generated by cylinders of Borel sets, and
$\mathbb{P}=\bigotimes_{k\in\Z}\mu$, where $\mu$ is a probability
measure on $\mathbb{T}$. The random
 variables $\theta_k$ on $(\Omega,\mathcal{F},\mathbb{P})$ are defined by
 \be\label{phases} \theta_k: \Omega \rightarrow \mathbb{T}, \ \
 \ \theta_k^\omega=\omega_{k}, \ \ \ k\in \Z.
\ee This ensures that the resulting operator
\be\label{U}U_\omega=D_\omega S\ee is unitary and ergodic with
respect to the $2$-shift in $\Omega$ \cite{bhj}. $U_\omega$ also
inherits the band structure of the original operator $S$ and has
the almost sure spectrum~\cite{j1}
$$ \Sigma =\exp(i\,\mbox{\em supp} \ \mu)\Sigma(t)=
\{e^{i\alpha}\Sigma(t)\,\,| \,\,\alpha\in\mbox{\em supp}\ \mu
\}.$$ Here supp$\,\mu$ denotes the support of the probability
measure $\mu$, defined as
$$ \text{supp}\,\mu:=\{a\;|\;\mu(a-\epsilon,a+\epsilon)>0\text{ for all
}\epsilon>0\}.$$
\par Solutions of the eigenvalue equation
$$U_{\omega}\psi=e^{i\lambda}\psi, \quad \psi=\sum_{k\in\Z}c_k e_k,$$
with $c_k\in\C$, $\lambda\in\C$, are characterized by the
relations
$$ \label{eigeneqn}  \begin{pmatrix}c_{2k+1} \\
c_{2k+2}\end{pmatrix}=
T(\theta_{2k}^\omega(\lambda),\theta_{2k+1}^\omega(\lambda))
\begin{pmatrix}c_{2k-1} \\ c_{2k}\end{pmatrix}, $$
for all $k\in\Z$, where the transfer matrices
$T:\mathbb{T}^2\rightarrow GL(2,\C)$ are defined by

\be \label{trensfer matrices} T(\theta, \eta) = \begin{pmatrix}
-e^{-i\eta} &
\frac{r}{t}\left(e^{i(\theta-\eta)} -e^{-i\eta}\right) \\
\frac{r}{t}\left(1-e^{-i\eta}\right) & -\frac{1}{t^2}\,e^{i\theta}
+\frac{r^2}{t^2}\left(1+e^{i(\theta-\eta)}-e^{-i\eta}\right)
\end{pmatrix}, \ee
and  the phases  by
\[\theta_k^\omega(\lambda)=\theta_k^\omega+\lambda.\]
  Note that det $T(\theta_{2k}^\omega(\lambda),\theta_{2k+1}^\omega(\lambda))=e^{i(\theta_{2k}^\omega-\theta_{2k+1}^\omega)}$ has modulus one and is independent of $\lambda$.

We have for any $n\in\N$
\begin{align}\label{cocycle} \begin{pmatrix}c_{2n-1} \cr
c_{2n}\end{pmatrix}&=T(\theta_{2(n-1)}^\omega(\lambda),\theta_{2(n-1)+1}^\omega(\lambda))\cdots
T(\theta_{0}^\omega(\lambda),\theta_{1}^\omega(\lambda))\begin{pmatrix}c_{-1}
\cr c_{0}\end{pmatrix}
 \equiv T_n (\omega,\lambda)\begin{pmatrix}c_{-1} \cr c_{0}\end{pmatrix} \nonumber \\
\begin{pmatrix}c_{-2n-1} \cr
c_{-2n}\end{pmatrix}&=T(\theta_{-2n}^\omega(\lambda),\theta_{-2n+1}^\omega(\lambda))^{-1}\cdots
T(\theta_{-2}^\omega(\lambda),\theta_{-1}^\omega(\lambda))^{-1}\begin{pmatrix}c_{-1}
\cr c_{0}\end{pmatrix} \equiv
T_{-n}(\omega,\lambda)\begin{pmatrix}c_{-1} \cr
c_{0}\end{pmatrix}.\nonumber \end{align} We also set
$T_0(\omega,\lambda)=\mathbb{I}$. \par As shown in~\cite{bhj}, for
any $\lambda\in\C $, the Lyapunov exponent $$\label{lyapu}
\gamma^{\pm}_{\omega}(\lm) =
\lim_{n\to\pm\infty}\frac{1}{|n|}\ln\|T_n(\omega,\lm)\|$$ almost
surely exists, has the same value for $k\to\infty$ and $k\to
-\infty$, and takes the deterministic value
\be\label{def.lyp.exp.}
\gm(\lambda)=\lim_{n\to\infty}{\dfrac{\E(\ln{||T_n(\omega,\lambda)||})}{n}}.\ee
A version of the Ishii-Pastur theorem suited to the present model
was proven in~\cite{bhj}.

\begin{Theorem}\label{IP} Let
$U_\omega$ be defined by (\ref{U}), (\ref{d}) and (\ref{phases})
and $\gamma (\lambda)$ by (\ref{def.lyp.exp.}). Then $$
\Sigma_{ac} \subseteq \overline{ \{ e^{i \lambda} \in S^1 ; \gamma
(\lambda)=0 \} }^{\mbox{ess}} \enspace . $$ \end{Theorem}

All norms on $GL(n,\C)$ being equivalent, we choose to work with
the row-sum norm for convenience. Thus, in what follows the norm
is the maximum row sum, i.e.\ for $A=(a_{ij})_{i,j=1}^{n}$,
$||A||=\max_{1\leq i\leq n}{\sum_{j=1}^{n}|a_{ij}|}$.
$\mathbf{P}(\C^2)$ denotes the projective space of $\C^2$, we
write $\bar{v}\in \mathbf{P}(\C^2)$ for the direction of $v\in
\C^2\setminus\{0\}$. The action of a $2\times 2$ matrix $A$ on
$\mathbf{P}(\C^2)$ is defined by $A\bar{v}=\overline{Av}$.

%%%%%%%%%%%%%%%%%%%%%%%%%%%%%%%%%%%%%%%%%%%%%%%%%%%%%%%%%%%%%%%%%%%%%%%%%%%%%%%%%%%%%%%%%%%%%%%%%%%%%%%%%%

\section{The Main Results}\label{results}

For a particular choice of the underlying distribution of the
random phases the unitary Anderson model, unlike the self-adjoint
one, exhibits two critical values of the spectral parameter, where
the Lyapunov exponent vanishes.
\begin{Theorem}\label{picase}
If supp$\,\mu=\{a,b\}$ and $|a-b|=\pi$, then
\begin{enumerate} [\upshape (i)]
\item $\gm(-a)=\gm(-b)=0$.\label{critical}
\item
 $\gm(\lambda)>0$, for all
$\lm\in\mathbb{T}\setminus\{-a,-b\}$.\label{pos.lyp.pi}
 \end{enumerate}
 \end{Theorem}

This is the only exceptional case. For all other choices of the
probability measure $\mu$ the Lyapunov exponent never vanishes.

\begin{Theorem}\label{pos.lyp}
If $\{a,b\}\subset$ supp$\,\mu$ such that $|a-b|\notin\{0,\pi\}$,
then for every $\lm\in\mathbb{T}$ we have $\gm(\lm)>0$.
 In particular, if supp$\,\mu$ contains at least three elements, then $\gm(\lm)>0$ for all $\lm\in\mathbb{T}$.
\end{Theorem}

Theorem~\ref{pos.lyp} is a generalization of the corresponding
results on the Lyapunov exponent previously proven in~\cite{bhj}
and~\cite{hjs}.

Theorem~\ref{picase}(i) will be proven in Section~\ref{crit},
while the proofs of Theorem~\ref{picase}(ii) and
Theorem~\ref{pos.lyp} are provided in
Section~\ref{sec:positivity}.

The fact that for all non-trivial probability measures $\mu$, the
set of critical quasi-energies contains at most two points
combined with Theorem~\ref{IP} gives the following immediate
corollary, concerning the almost sure absolutely continuous
spectrum of $U_\omega$.

\begin{Corollary}\label{ac spectrum}
For any non-trivial distribution $\mu$ of the i.i.d.\ random
phases, we have
$$\Sigma_{ac}=\emptyset. $$
\end{Corollary}
%%%%%%%%%%%%%%%%%%%%%%%%%%%%%%%%%%%%%%%%%%%%%%%%%%%%%%%%%%%%%%%%%%%%%%%%%%%%%%%%%%%%%%%%%%%%%%%%%%%%%%%%%

%%%%%%%%%%%%%%%%%%%%%%%%%%%%%%%%%%%%%%%%%%%%%%%%%%%%%%%%%%%%%%%%%%%%%%%%%%%%%%%%%%%%%
\section{Critical Quasi-energies} \label{crit}

In this section we prove Theorem~\ref{picase}(i), i.e.\ that a
Bernoulli measure $\mu$ with diametrically opposed masses $a$, $b$
indeed gives rise to two critical quasi-energies at $\lm=-a$,
$\lm=-b$. Denote $\mu(a) = p \in (0,1)$ and $\mu(b) = q = 1-p$.

For $\lm=-a$, the i.i.d.\
 random matrices
 $T(\theta_{2k}^\omega(\lambda),\theta_{2k+1}^\omega(\lambda))$
  take only the following values with non-zero probabilities,
\be\label{old matrices}
 T(\theta_{2k}^\omega(\lambda),\theta_{2k+1}^\omega(\lambda)) =
 \begin{cases}
 T(0,0)=-I,& \text{with probability $p^2$}
 \\ T(\pi,\pi)={
\begin{pmatrix}
   1&2r/t\\
2r/t&(3r^2+1)/t^2\\
    \end{pmatrix}},& \text{with probability $q^2$}
 \\ T(\pi,0)={
\begin{pmatrix}
   -1&-2r/t\\
0&1\\
    \end{pmatrix}},& \text{with probability $pq$}
 \\T(0,\pi)= {
\begin{pmatrix}
   1&0\\
2r/t&-1\\
    \end{pmatrix}},& \text{with probability $pq$.}
 \end{cases}
 \ee
 \par The latter matrices take much simpler forms when
 represented with respect to the basis $\{{ \begin{pmatrix}
   1\\
(r+1)/t\\
    \end{pmatrix}},{ \begin{pmatrix}
   1\\
(r-1)/t\\
    \end{pmatrix}}\}$ of $\C^2$. Hence, we define the matrices
    $A(\theta,\eta)$ as
    \[A(\theta,\eta):={ \begin{pmatrix}
   1&1\\
(r+1)/t&(r-1)/t\\
    \end{pmatrix}}^{-1}T(\theta,\eta) { \begin{pmatrix}
   1&1\\
(r+1)/t&(r-1)/t\\
    \end{pmatrix}}.\]
It follows that
\begin{eqnarray} \label{new matrices}
 A(0,0)=-I, & & A(\pi,\pi)={\begin{pmatrix}
   \rho&0\\
0&1/\rho \end{pmatrix}}\nonumber \\
    A(\pi,0)={ \begin{pmatrix}
   0&-1/\rho\\-\rho&0 \end{pmatrix}}, & & A(0,\pi)={ \begin{pmatrix}
   0&1\\1&0 \end{pmatrix}},
\end{eqnarray}
    with $\rho:=(r+1)^2/t^2>1$.
    \par A straightforward calculation shows that
\be\label{before projection}
    \gm(\lambda)=\lim_{n\to\infty}{\dfrac{\E(\ln{||\Lambda_n(\omega,\lambda)||})}{n}},
\ee where
$\Lambda_n(\omega,\lambda)=\Pi_{k=1}^nA(\theta_{2k}^\omega(\lambda),\theta_{2k+1}^\omega(\lambda))$.
%%%%%%%%%%%%%%%%%%%%%%%%%%%%%%%%%%%%%%%%%%%%%%%%%%%%%%%%%%%%%%%%%%%%%%%%%%%%%%
 In order to simplify the notation, we will suppress the $\omega$
dependence of various quantities for the remainder of
the section. \par Let $u_0:={\begin{pmatrix} 1\\ 1\\
\end{pmatrix}}$ and
$u_n(\lambda)={\begin{pmatrix}
u_{n,1}(\lambda)\\ u_{n,2}(\lambda)\\
\end{pmatrix}} := \Lambda_n(\lambda)u_0$.
 \begin{Lemma}\label{projection}
 If $x_n=\dfrac{\ln|u_{n,1}(\lm)|}{\ln\rho}$ for $n\geq0$, then
\be\label{after projection1}
\gm(-a)=\ln\rho\lim_{n\to\infty}{\dfrac{1}{n}\E(|x_n|)}. \ee
 \end{Lemma}
 \begin{proof} From (\ref{new matrices}) it follows
that with probability one there are $x,y\in\R$ with $xy=1$ and
either $\Lambda_n(-a)={\begin{pmatrix} x&0\\ 0&y\\
\end{pmatrix}}$ or $\Lambda_n(-a)={\begin{pmatrix} 0&x\\ y&0\\
\end{pmatrix}}$. In both cases it follows readily that \[
||\Lambda_n(-a)||= ||\Lambda_n(-a)u_0||_\infty,\]
 where $||\cdot||_\infty$ denotes the max-norm on $\C^2$. This
 implies that
 \be\label{after projection}
 \gm(-a)=\lim_{n\to\infty}{\dfrac{1}{n}\E(\ln{||\Lambda_{n}(-a)u_0||_\infty})}.\ee
 Furthermore we see from the specific form of $\Lambda_n(-a)$
 that $\Lambda_n(-a)u_0={\begin{pmatrix}
u_{n,1}(-a)\\
1/u_{n,1}(-a)\\
\end{pmatrix}}$. Therefore
\begin{align*}
\ln||\Lambda_n(-a)u_0||_\infty&=\ln\max(|u_{n,1}(-a)|,\dfrac{1}{|u_{n,1}(-a)|})
\\&=|\ln|u_{n,1}(-a)||.
\end{align*}
 The required result then follows from (\ref{after projection}) and the definition of $x_n$.
\end{proof}
%%%%%%%%%%%%%%%%%%%%%%%%%%%%%%%%%%%%%%%%%%%%%%%%%%%%%%%%%%%%%%%%%%%%%%%%%%%%%%
The following lemma is devoted to the necessary analysis of the
random sequence $x_n$.
\begin{Lemma}
$(x_n)_{n\geq0}$ is an integer-valued Markov chain with $x_0=0$
and transition probabilities \begin{eqnarray} \label{markov chain}
 \PP(x_{n+1}=x_n)=p^2, & & \PP(x_{n+1}=x_n+1)=q^2,\notag
\\\PP(x_{n+1}=-x_n)=pq, & & \PP(x_{n+1}=-(x_n+1))=pq.
\end{eqnarray}
\end{Lemma}
\begin{proof}
Clearly, $x_0=0$. Let
$A(n+1,-a):=A(\theta_{2(n+1)}-a,\theta_{2(n+1)+1}-a))$. In the
case $A(n+1,-a)=-I$, we have $|u_{n+1,1}(-a)|=|u_{n,1}(-a)|$, i.e.
$x_{n+1}=x_n$. If $A(n+1,-a)={\begin{pmatrix}
\rho&0\\
0&1/\rho\\
\end{pmatrix}}$, then $|u_{n+1,1}(-a)|=\rho|u_{n,1}(-a)|$ and
$x_{n+1}=x_n+1$. Similarly, $A(n+1,-a)={\begin{pmatrix}
0&1\\
1&0\\
\end{pmatrix}}$ implies that $|u_{n+1,1}(-a)|=|u_{n,2}(-a)|=1/|u_{n,1}(-a)|$
and thus $x_{n+1}=-x_n$. Finally, $A(n+1,-a)={\begin{pmatrix}
0&-1/\rho\\
-\rho&0\\
\end{pmatrix}}$ gives
$|u_{n+1,1}(-a)|=\dfrac{1}{\rho}|u_{n,2}(-a)|=\dfrac{1}{\rho|u_{n,1}(-a)|}$,
i.e. $x_{n+1}=-x_n-1$. Thus $x_{n+1}$ is determined by $x_n$ and
$A(n+1,-a)$. The transition probabilities follow from (\ref{old
matrices}) and (\ref{new matrices}).
\end{proof}
%%%%%%%%%%%%%%%%%%%%%%%%%%%%%%%%%%%%%%%%%%%%%%%%%%%%%%%%%%%%%%

\begin{Lemma}\label{expect}
  As $n\to\infty$,
\be \label{asymp} \dfrac{\E(x_n^2)}{n}\to \frac{q}{2p}. \ee
  In particular, we have that for all $n$,
   \[\E(|x_n|)\leq C n^{1/2}.\]
  \end{Lemma}
  \begin{proof}
  Let $\alpha=q-p$, we denote by $\E(x|y)$ the conditional
 expectation of $x$ given $y$. It follows that
 \begin{align*}
\E(x_n|x_{n-1})=\alpha^2x_{n-1}+q\alpha.
 \end{align*}
 Since $\E(x_n)=\E(\E(x_n|x_{n-1}))$, we get that
  \[\E(x_n)=\alpha^2\E(x_{n-1})+q\alpha\] and, iterating,
 \begin{align*}
 \E(x_n)&=\alpha^{2n}\E(x_0)+q\alpha\dfrac{1-\alpha^{2n}}{1-\alpha^2}.
 %\\&=q\alpha\dfrac{1-\alpha^{2(n-1)}}{1-\alpha^2},&&\text{since
 %$\E(x_0)=0$}.
 \end{align*}
Similarly, since
\[\E(x_n^2|x_{n-1})=x_{n-1}^2+2qx_{n-1}+q,\] we have that
\[\E(x_n^2)=\E(x_{n-1}^2)+2q\E(x_{n-1})+q.\]
Another induction gives that, for all $n\geq3$,
\begin{eqnarray*}
\E(x_n^2) & = &
\E(x_0^2)+2q\dfrac{1-\alpha^{2n}}{1-\alpha^2}\E(x_0)+nq(1+\dfrac{2q\alpha}{1-\alpha^2})\\
& &
\mbox{}+2q^2\alpha[1-\dfrac{1}{1-\alpha^2}(2+\alpha^4\dfrac{1-\alpha^{2(n-2)}}{1-\alpha^2})].
\end{eqnarray*}
Since $|\alpha|<1$ and $\E(x_0)=\E(x_0^2)=0$ and
$q(1+\dfrac{2q\alpha}{1-\alpha^2})=q/2p$, we get (\ref{asymp}).
Using that $\E(|x_n|)\leq(\E(x_n^2))^{1/2}$, in turn, proves the
second assertion and finishes the proof.
  \end{proof}

Lemma~\ref{expect} is, in fact, a consequence of general
extensions of the Central Limit Theorem used in the study of
dynamical systems, e.g.\ Section~A.4 of \cite{cd}. We include the
previous elementary proof for the convenience of the reader.

The main result of this section now follows immediately.

 \begin{proof}[\textbf{Proof of Theorem~\ref{picase}(\ref{critical}})]
The fact that $\gm(-a)=0$ follows  directly from (\ref{after
projection1}) and Lemma~\ref{expect}. The proof of
 $\gm(-b)=0$  is identical.
 \end{proof}

%%%%%%%%%%%%%%%%%%%%%%%%%%%%%%%%%%%%%%%%%%%%%%%%%%%%%%%%%%%%%

\section{Positivity of the Lyapunov Exponent} \label{sec:positivity}

In this section we show that, except for the two critical energies
discussed above, the Lyapunov exponent (\ref{def.lyp.exp.}) is
positive. This constitutes the contents of part~(\ref{pos.lyp.pi})
of Theorem~\ref{picase} and of Theorem~\ref{pos.lyp}. For each
$\lambda \in \mathbb{T}$, the random variables $\theta_0^{\omega}$
and $\theta_1^{\omega}$ induce a measure on $GL(2,\C)$ through
$T(\theta_0^{\omega}+\lambda, \theta_1^{\omega}+\lambda)$. Denote
the smallest closed subgroup of $GL(2,\C)$ generated by the
support of this measure by $G_{\lambda,\mu}$. Thus
$G_{\lambda,\mu}$ is generated by the matrices $T(\theta,\eta)$,
defined in (\ref{trensfer matrices}), where $\theta$ and $\eta$
vary in $\lambda + \mbox{supp}\,\mu$.

F\"urstenberg's Theorem~\cite{bl} states that if $G_{\lambda,\mu}$
is non-compact and strongly irreducible, then \begin{align*}
\gm(\lambda)=\lim_{n\to\infty}{\dfrac{\E(\ln{||T_n(\omega,\lambda)||})}{n}}>0.
\end{align*} \par The proof that $G_{\lambda,\mu}$ is non-compact
for all values of $\lm\in\mathbb{T}$ was given in~\cite{hjs} and
holds for any non-trivial probability distribution $\mu$. For
completeness we repeat the proof here.
\begin{Lemma}\label{non-compact}
$G_{\lambda,\mu}$ is non-compact.
\end{Lemma}
\begin{proof}
Let $\theta$ and $\eta$ be on the torus, $\theta \not=\eta$, and
let $x:= e^{-i\theta}$, $z:= e^{-i\eta}$. Let $G(\theta,\eta)$ be
the closed group generated by $T(\theta,\theta)$, $T(\eta,\eta)$,
$T(\theta,\eta)$, and $T(\eta,\theta)$. Define \be \label{oper.D}
D:= T(\theta,\theta)
T(\theta,\eta)^{-1} = \left( \begin{array}{ll} x\bar{z} & 0 \\
\frac{r}{t}(x\bar{z}-1) & 1 \end{array} \right) \in
G(\theta,\eta),\ee
 \be \label{oper.E} E := T(\eta,\theta)^{-1} T(\theta,\theta) =
\left(
\begin{array}{ll} 1 & \frac{r}{t} (1-\bar{x}z) \\ 0 & \bar{x}z
\end{array} \right) \in G(\theta,\eta),\ee
 \be \label{oper.L} L := DE =
\left( \begin{array}{ll} x\bar{z} & \frac{r}{t} (x\bar{z}-1) \\
\frac{r}{t} (x\bar{z}-1) & \bar{x}z-\frac{r^2}{t^2} |x\bar{z}-1|^2
\end{array} \right) \in G(\theta,\eta),\ee
 \be \label{oper.J} J := ED =
\left( \begin{array}{ll} x\bar{z} - \frac{r^2}{t^2} |\bar{x}z-1|^2
& \frac{r}{t}(1-\bar{x}z) \\ \frac{r}{t}(1-\bar{x}z) & \bar{x}z
\end{array} \right) \in G(\theta,\eta).\ee

Note that $\det L = \det J = 1$ and that $J^{-1} = L^*$. Thus we
get the self-adjoint element $K := J^{-1}L$ of $G(\theta,\eta)$.
In fact, $K$ is positive definite and $\det K=1$. More calculation
shows that \begin{eqnarray*} \mbox{tr}\,K & = & 1 +
\frac{2r^2}{t^2} |x\bar{z}-1|^2 + \left|x\bar{z}- \frac{r^2}{t^2}
|x\bar{z}-1|^2\right|^2 \\ & = & 2 + \frac{r^2}{t^4}
|x\bar{z}-1|^4.
\end{eqnarray*}

As $\theta \not= \eta$ and therefore $x\bar{z} \not= 1$ we
conclude that tr$\,K >2$. Positivity of $K$ implies that it has an
eigenvalue strictly bigger than $1$. Thus, containing all powers
of $K$, the group $G(\theta,\eta)$ is non-compact. In particular,
with $\theta=\lambda+a$ and $\eta=\lambda+b$, we see that
$G_{\lambda,\mu}$ is non compact.
\end{proof}

It remains to prove strong irreducibility under the assumptions of
Theorem~\ref{picase}(\ref{pos.lyp.pi}) as well as under those of
Theorem~\ref{pos.lyp}. Under the already established
non-compactness of $G_{\lambda,\mu}$, strong irreducibility of
$G_{\lambda,\mu}$ is equivalent to
\begin{equation}\label{strong irr}
\#\{g\bar{v}:g\in G_{\lambda,\mu}\}\geq 3 \quad \mbox{for all
$\bar{v}\in\mathbf{P}(\C^2)$},
\end{equation}
see~\cite{bl}. We first use this fact to prove that for
supp$\,\mu=\{a,b\}$ and $|a-b|=\pi$, $-a$ and $-b$ are the only
critical quasi-energies.
%%%%%%%%%%%%%%%%%%%%%%%%%%%%%%%%%%%%%%%%%%%%%%%%%%%%%%%%%%%%%%%%%%%%%%%%%%%%%%%%%%%%%%%%
\begin{proof}[\textbf{Proof of Theorem~\ref{picase}(\ref{pos.lyp.pi})}] Let
$\theta=a+\lm$, $\eta=b+\lm$, with
$\lm\in\mathbb{T}\backslash\{-a,-b\}$. In the terminology
introduced above, the condition that $|a-b|= \pi$ can be written
as $-x=z\notin\{-1,1\}$. Since $x=-z$, the operator $L\in
G_{\lambda,\mu}$ defined by (\ref{oper.L}) takes the form
    \[L= {
\begin{pmatrix}
   -1&\dfrac{-2r}{t}\\
\dfrac{-2r}{t}&-1-\dfrac{4r^2}{t^2}\\
    \end{pmatrix}}.\]
 As det$\,L=1$ and $|$tr$L|>2$, $L$ is hyperbolic, hence iterations of $L$ map any direction in
 $\mathbf{P}(\C^2)$ to infinitely many directions, except when
 $\overline{v}$ coincides with the direction of one of its eigenvectors,
 given by $v_+={
\begin{pmatrix}
  1\\
\dfrac{r+1}{t}\\
   \end{pmatrix}}$, $v_-={
\begin{pmatrix}
  1 \\
\dfrac{r-1}{t}\\
   \end{pmatrix}}$.
\par  Next we prove that even for the eigenvectors of $L$, we have
that $\#\{g\bar{v}_{\underset{-}{+}}:g\in G_{\lambda,\mu}\}\geq
3$.
\par Under the current conditions, the transfer matrices take
    the form
 \[T(\theta,\theta)= {
\begin{pmatrix}
   -x&\dfrac{r}{t}(1-x)\\
\dfrac{r}{t}(1-x)&\dfrac{r^2}{t^2}(2-x)-\dfrac{1}{t^2}\bar{x}\\
    \end{pmatrix}},\]
\[T(\theta,\eta)= { \begin{pmatrix}
   x&\dfrac{r}{t}(-1+x)\\
\dfrac{r}{t}(1+x)&\dfrac{r^2}{t^2}x-\dfrac{1}{t^2}\bar{x}\\
    \end{pmatrix}},\]
\[T(\eta,\theta)= { \begin{pmatrix}
   -x&\dfrac{-r}{t}(1+x)\\
\dfrac{r}{t}(1-x)&\dfrac{-r^2}{t^2}x+\dfrac{1}{t^2}\bar{x}\\
    \end{pmatrix}},\]
    \[T(\eta,\eta)= {
\begin{pmatrix}
   x&\dfrac{r}{t}(1+x)\\
\dfrac{r}{t}(1+x)&\dfrac{r^2}{t^2}(2+x)+\dfrac{1}{t^2}\bar{x}\\
    \end{pmatrix}}.\]
    Therefore, we have that
\[ T(\theta,\theta)v _+= {\begin{pmatrix}
 -\dfrac{r+1}{t^2}(r-x)\\
 -\dfrac{r+1}{t^3}(rx+\bar{x})+\dfrac{r(r+1)^2}{t^3}  \end{pmatrix} }, \]
\[ T(\theta,\eta)v _+= {\begin{pmatrix}
 -\dfrac{r+1}{t^2}(r-x)\\
 \dfrac{r+1}{t^3}(rx-\bar{x})+\dfrac{r}{t}  \end{pmatrix} }, \]
    \[ T(\eta,\theta)v _+= {\begin{pmatrix}
 -\dfrac{r+1}{t^2}(r+x)\\
 -\dfrac{r+1}{t^3}(rx-\bar{x})+\dfrac{r}{t}  \end{pmatrix} }. \]
\par A simple calculation shows that
  $\overline{T(\theta,\theta)v_+}=\overline{T(\theta,\eta)v_+}$  only if  $x=r$.
  Similarly,
$\overline{T(\theta,\eta)v}=\overline{T(\eta,\theta)v}$  only if
$x\in\{-1,1\}$, while assuming that
$\overline{T(\theta,\theta)v_+}=\overline{T(\eta,\theta)v_+}$ is
equivalent to $(r+1)t^2=0$. All these cases are excluded by the
assumptions of Theorem~\ref{picase}(\ref{pos.lyp.pi}). Therefore,
we conclude that $\overline{T(\theta,\theta)v_+}$,
$\overline{T(\theta,\eta)v_+}$, and $\overline{T(\eta,\theta)v_+}$
are all different. In a similar way one treats $v_-$. We thus have
proven that
\[\#\{g\bar{v}:g\in G_{\lambda,\mu}\}\geq 3 \quad \mbox{for all
$\bar{v}\in\mathbf{P}(\C^2)$}.\] Combining this with
Lemma~\ref{non-compact}, F\"urstenberg's Theorem gives the
required assertion.
\end{proof}
 %%%%%%%%%%%%%%%%%%%%%%%%%%%%%%%%%%%%%%%%%%%%%%%%%
\begin{proof}[\textbf{Proof of Theorem \ref{pos.lyp}}]
 Again by F\"urstenberg's Theorem and Lemma~\ref{non-compact},
  proving that $\gm(\lm)>0$ for all $\lm\in\C$ for the case that the support of
  $\mu$ contains two points of $\mathbb{T}$ that are not diametrically
  opposed, is reduced to checking condition (\ref{strong irr}). Each element
  of the projective space
  $\mathbf{P}(\C^2)$ is of the form $\bar{v}$ with $v=\begin{pmatrix}
   0\\
1\\
    \end{pmatrix}$ or $v=\begin{pmatrix}
   1\\
\alpha\\
    \end{pmatrix}$, for some $\alpha\in\C$.
 In terms of $x,z$, introduced above, the condition that
$|a-b|\notin\{0,\pi\}$
  can be written as  $x\bar{z}\notin\{-1,1\}$.
 \par \textbf{Case I:} Let $v= {
\begin{pmatrix}
   0\\
1\\
    \end{pmatrix}}$. The action of the operator $E$, defined in (\ref{oper.E}), on $\bar{v}$  has the direction of $ {
\begin{pmatrix}
   \dfrac{r}{t}(x\bar{z}-1)\\
1\\
    \end{pmatrix}}$, while $E^2\bar{v}$ has the direction of ${
\begin{pmatrix}
   \dfrac{r}{t}((x\bar{z})^2-1)\\
1\\
    \end{pmatrix}}$. Thus $\{I,E,E^2\}\subset G_{\lambda,\mu}$
    maps $\bar{v}$ into three different elements in $\mathbf{P}(\C^2)$.
 \par \textbf{Case II:}  Let  $v= {
\begin{pmatrix}
   1\\
\alpha\\
    \end{pmatrix}}$, with $\alpha\in \C$. Acting on $\bar{v}$ with the operator $D$ from (\ref{oper.D})
    results in the direction of ${
\begin{pmatrix}
  1 \\
\dfrac{r}{t}(1-\bar{x}z)+\alpha\bar{x}z\\
    \end{pmatrix}}$, while $E\bar{v}$ has the direction of ${
\begin{pmatrix}
  1 \\
\dfrac{\alpha\bar{x}z}{1+\alpha r/t(1-\bar{x}z)}\\
    \end{pmatrix}}$.
      \par Defining the map $F$ such that
    $F:c\mapsto\dfrac{r}{t}(1-\bar{x}z)+c\bar{x}z$, one sees that $F$ has a single fixed
    point at $c=r/t$. Since the second iteration $F^2:c\mapsto\dfrac{r}{t}(1-(\bar{x}z)^2)+c(\bar{x}z)^2$, has
    the same value $r/t$ as its only fixed point, we deduce that $\{c,F(c),F^2(c)\}$ are pairwise different except when $c=r/t$.
     Thus iterations of the operator
$D$ take $\bar{v}$
     into at least three different directions, unless $\alpha=r/t$.
      \par On the other hand, the map
      $H:c\mapsto\dfrac{c\bar{x}z}{1+cr/t(1-\bar{x}z)}$ has fixed points $0,-t/r$, which are also the fixed points of $H^2:c\mapsto\dfrac{c(\bar{x}z)^2}{1+cr/t(1-(\bar{x}z)^2)}$.
     In particular, $\{I,E,E^2\}\subset G_{\lambda,\mu}$
       map the direction vector ${
\begin{pmatrix}
  1 \\
r/t\\
    \end{pmatrix}}$ to three different elements in
    $\mathbf{P}(\C^2)$. This proves the required condition for
    strong irreducibility of $G_{\lambda,\mu}$ and once more the
    result of Lemma \ref{non-compact} and F\"urstenberg's Theorem
    finish the proof.
\end{proof}
%%%%%%%%%%%%%%%%%%%%%%%%%%%%%%%%%%%%%%%%%%%%%%%%%%%%%%%%%%%%%%%%%%%%%%%%%%%%%%%%%%%%%%%%%%%%%%%%%%%%%%%%%%%%%%%%%%%%%%%%%%%%%%%%%%%%%%%%%%%%%%%%%%%%%%%%%%%%
%%%%%%%%%%%%%%%%%%%%%%%%%%%%%%%%%%%%%%%%%%%%%%%%%%%%%%%%%%%%%%%%%%%%%%%%%%%%%%%%%%%%%%%%%%%%%%%%%%%%%%%%%%%%%%%%%%%%%%%%%%%%%%%%%%%%%%%%%%%%%%%%%%%%%%%%%%%%%%%%%%%%
%%%%%%%%%%%%%%%%%%%%%%%%%%%%%%%%%%%%%%%%%%%%%%%%%%%%%%%%%%%%%%%%%%%%%%%%%%%%%%%%%%%%%%%%%%%%%%%%%%%%%%%%%%%%%%%%%%%%%%%%%%%%%%%%%%%%%%%%%%%%%%%%%%%%%
\section{A Unitary Dimer Model} \label{dimer}

In this section we study a unitary version of the Dimer model,
which is obtained from the Anderson model (\ref{U}) by doubling up
the random phases. More rigorously, let a probability measure
$\mu$ on $\mathbb{T}$ be given, define $g:\T\to\T^2$ as
$g(\theta)=(\theta,\theta)$ and let $\tilde{\mu}$ be the
probability measure supported on the diagonal of $\mathbb{T}^2$
induced by $\mu$ through $g$: $\tilde{\mu}(\tilde{B}) =
\mu(g^{-1}(\tilde{B}))$ for Borel sets $\tilde{B}$ in
$\mathbb{T}^2$. We introduce the probability space
$(\tilde{\Omega},\tilde{\mathcal{F}},\tilde{\mathbb{P}})$, where
$\tilde{\Omega}$ is identified with $(\mathbb{T}^2)^\Z$,
$\tilde{\mathcal{F}}$ is the $\sigma$-algebra generated by
cylinders of Borel sets in $\mathbb{T}^2$ and
$\tilde{\mathbb{P}}=\bigotimes_{k\in\Z}\tilde{\mu}$.

For $\omega = (\omega_k)_{k\in\Z} \in \tilde{\Omega}$, the random
phases $\theta_n^{\omega}$, $n\in\Z$, used in (\ref{d}) and
(\ref{U}) to define $U_{\omega}$ are now chosen as \begin{align*}
(\theta_{2k},\theta_{2k+1})=\omega_{k}, \quad k\in \Z.\end{align*}

 For this model we again prove that the almost sure absolutely continuous spectrum is empty.
 We also show that the case of a Bernoulli measure
\be\label{bern.meas} \mu=p\delta_a+q\delta_b, \quad p+q=1, \quad
a,b\in\mathbb{T}, \quad a\not= b,\ee
  gives rise to
additional critical quasi-energies, as this is the case of least
randomness.

The following theorem states that for any non-trivial distribution
$\mu$ on $\T$, i.e. $\{a,b\}\subset$ supp$\,\mu$ for $a\neq b$, the
Lyapunov exponent is positive for all but a finite set of
quasi-energies, given by
 \begin{align*}
M:=\{-a,-b\}\cup&\big(M_a\cap M_b\big), \end{align*}
 where
 \begin{align*}
 \hspace{-1cm}M_a&=\{\arccos(r^2)-a,2\pi-\arccos(r^2)-a,\arccos(r^2-t^2)-a,2\pi-\arccos(r^2-t^2)-a\},
\\\hspace{-1cm}M_b&=\{\arccos(r^2)-b,2\pi-\arccos(r^2)-b,\arccos(r^2-t^2)-b,2\pi-\arccos(r^2-t^2)-b\}.
\end{align*}
 An immediate consequence is that the almost sure absolutely continuous spectrum
 of these operators is trivial.

\begin{Theorem}\label{t:d lyp}
If $\{a,b\}\subset$ supp$\,\mu$, then for all
$\lm\in\mathbb{T}\backslash M$, the Lyapunov exponent $\gm(\lambda)$
is strictly positive. In particular, $\Sigma_{ac}=\emptyset$.
\end{Theorem}

As before, we will use F\"urstenberg's Theorem~\cite{bl} to prove
positivity of Lyapunov exponents. Let $\widetilde{G}_{\lambda,\mu}$
be the closed group corresponding to $G_{\lambda,\mu}$ from the
previous section.

We will show that $\widetilde{G}_{\lambda,\mu}$ is both non-compact
and strongly irreducible for all $\lm$ outside of $M$. As both of
these properties carry over to larger groups, we may assume for the
rest of the proof of Theorem~\ref{t:d lyp} that supp$\,\mu =
\{a,b\}$. Thus $\widetilde{G}_{\lambda,\mu}$ is generated by just
two matrices, $T(\theta,\theta)$ and $T(\eta,\eta)$, where
 $\theta=a+\lambda$, $\eta=b+\lambda$.

%%%%%%%%%%%%%%%%%%%%%%%%%%%%%%%%%%%%%%%%%%%%%%%%%%%%%%%%%%%%%%%%%%%%%%%%%%%%%%%%%%%%%%%%%%%%%%%
 \par In order to prove Theorem~\ref{t:d lyp}, we start by mapping the problem into a somewhat simpler form.
 In order to simplify the notation, we again let $x:= e^{-i\theta}$, $z:= e^{-i\eta}$, and let
$\rho,\rho_1$ be the two eigenvalues of $T(\theta,\theta)$. Since
tr$\,T(\theta,\theta)=\dfrac{2r^2}{t^2}-\dfrac{1}{t^2}(x+\bar{x})$,
and $\det T(\theta,\theta)=\det T(\eta,\eta)=1$, we have the
following cases \be\label{e:eigenvalues}
\begin{cases}
\rho=\overline{\rho_1},\,|\rho|=1,& \text{when
$|$tr$\,T(\theta,\theta)|<2$}
 \\\rho=\dfrac{1}{\rho_1}>1,& \text{when $|$tr$\,T(\theta,\theta)|>2$}
 \\\rho=\rho_1, \,\rho^2=1,& \text{when $|$tr$\,T(\theta,\theta)|=2$}.
 \end{cases}
 \ee
 This allows us to introduce the transformation $N$ given by
 \be\notag
 N=\begin{pmatrix}
   \dfrac{r}{t}(1-x)&x+\rho\\
x+\rho&-\dfrac{r}{t}(1-x)\\
    \end{pmatrix}.
 \ee
Using that $(x+\rho)(x+\rho_1)=-\dfrac{r^2}{t^2}(1-x)^2$, we
deduce that $\det N=(x+\rho)(\rho_1-\rho)$. Therefore, $N$ is
invertible as long as $|$tr$\,T(\theta,\theta)|\neq2$. Moreover,
\be\label{E-matrix}
 E=NT(\theta,\theta)N^{-1}=\begin{pmatrix}
   \rho&0\\
0&1/\rho\\
    \end{pmatrix}.
\ee
\par A short calculation shows that the elements of
$F=NT(\eta,\eta)N^{-1}$ are given by
\begin{align}\label{F-matrix}
F_{11}&=\dfrac{1}{t^2(\rho-1/\rho)}[2r^2(1+\rho)-(z\bar{x}+\bar{z}x)-\rho(z+\bar{z})]\notag,
\\F_{12}&=F_{21}=\dfrac{2ir}{t^3(\rho-1/\rho)}[\Im(x)-\Im(z)+\Im(z\overline{x})]\notag,
\\F_{22}&=\dfrac{-1}{t^2(\rho-1/\rho)}[2r^2(1+\dfrac{1}{\rho})-(z\bar{x}+\bar{z}x)-\dfrac{1}{\rho}(z+\bar{z})].
\end{align}
 Notice that since $\theta\neq\eta$, $F_{12}=0$ if and only if either $\eta=0$ or $\theta=0$.
\par Since proving non-compactness and strong irreducibility of
 $\widetilde{G}_{\lambda,\mu}$ is equivalent to proving the same properties for the
 group $\widetilde{H}_{\lambda,\mu}$
 generated by the matrices $E$, $F$, we will use the latter, somewhat simpler
 matrices whenever it helps simplifying the proofs.
%%%%%%%%%%%%%%%%%%%%%%%%%%%%%%%%%%%%%%%%%%%%%%%%%%%%%%%%%%%%%%%%%%%%%%%%%%%%%%%%%%%%%%%%%%%%%%%%%%%%%%%%%%%
 \begin{Lemma}\label{t:d noncomp}
 For all $\lm\in \mathbb{T}\backslash\{-a,-b\}$, the group $\widetilde{G}_{\lambda,\mu}$ is
 non-compact.
\end{Lemma}
\begin{proof}
Since $\lm\neq -a$, we have that $\theta\neq0$ and thus
$trT(\theta,\theta)\neq-2$. Therefore, the preceding discussion
suggests the proof should be divided into the following cases;
 \par \textbf{Case I:} tr $T(\theta,\theta)=2$. By (\ref{e:eigenvalues})  we have $\rho=\rho_1=1$ and
  since by definition $T(\theta,\theta)\neq I$ it follows that there exists a non-singular matrix $R$ such
 that
 \[RT(\theta,\theta)R^{-1}=\begin{pmatrix}
   \rho&1\\
0&\rho\\
    \end{pmatrix}.\]
    Since $||[RT(\theta,\theta)R^{-1}]^n||$ grows with $n$, the
    group generated by
    $RT(\theta,\theta)R^{-1}$, $RT(\eta,\eta)R^{-1}$ is non-compact,
    which implies that $G_{\lambda,\mu}$ is non-compact.
\par \textbf{Case II:} $|$tr$\,T(\theta,\theta)|>2$, again by
(\ref{e:eigenvalues}), $T(\theta,\theta)$ has an eigenvalue $\rho>
1$ which gives the required result.
\par \textbf{Case III:} $|$tr$\,T(\theta,\theta)|<2$. In this case
equations (\ref{e:eigenvalues}), (\ref{E-matrix})  give
\[
E=\begin{pmatrix}
   e^{iy}&0\\
0&e^{-iy}\\
    \end{pmatrix},\]
with $y=\arccos(\dfrac{r^2}{t^2}-\dfrac{1}{2t^2}(x+\bar{x}))$, and
$y\in(0,\pi)$.  Equations (\ref{F-matrix}) lead to
\[ F=\begin{pmatrix}
   \al e^{ic}&\beta\\
\beta&\al e^{-ic}\\
    \end{pmatrix}, \quad \alpha\geq0, \;\beta\in\R, \;c\in\mathbb{T}.\]
Using that $\det F=1$, it follows that $\alpha>0$.
\par Now we follow a strategy outlined in~\cite{dss} to show  that there
exists a sequence of elements in $\widetilde{H}_{\lambda,\mu}$
with unbounded norms. In order to do so, we  note that any element
of $\mathbf{P}(\C^2)$ can be written in the form
\[e_{(u,v)}={
\begin{pmatrix}
   e^{iu}\cos(v)\\
e^{-iu}\sin(v)\\
    \end{pmatrix}}, \quad (u,v)\in[0,\pi)\times[0,\pi).\]
Therefore, for any element $e_{(u,v)}$ of $\mathbf P(\C^2)$ we
have
\[Fe_{(u,v)}={ \begin{pmatrix}
  \al e^{i(c+u)}\cos(v)+\beta e^{-iu}\sin(v)\\
\al e^{-i(c+u)}\sin(v)+\beta e^{iu}\cos(v)\\
    \end{pmatrix}}.\]
 Using that $\al^2-\beta^2=1$, we get
\[||Fe_{(u,v)}||^2-1=2\beta^2+4\al\beta\cos(2u+c)\cos(v)\sin(v).
\]
\par Since
$\beta=\dfrac{r}{t^3}\dfrac{\Im(x)-\Im(z)+\Im(z\overline{x})}{\Im(\rho)}$
with distinct $x,z$ and neither equals $1$ under the current
assumptions, we see that $\beta\neq 0$.
\par In the case  $\beta>0$:  If $\cos(v)\sin(v)=0$, then $||Fe_{(u,v)}||^2-1>\beta^2$
for all $u\in[0,\pi)$. While for $\cos(v)\sin(v)>0$,
$||Fe_{(u,v)}||^2-1>\beta^2$ is equivalent to
\[\cos(2u+c)>\dfrac{-\beta}{4\al\cos(v)\sin(v)}.\]
In particular, the condition $\cos(2u+c)>\dfrac{-\beta}{4\alpha}$
guarantees that $||Fe_{(u,v)}||^2>1+\beta^2$. Defining
$K_+:=\{u\in[0,\pi):\cos(2u+c)>\dfrac{-\beta}{4\alpha}\}$, we see
that $|K_+|>\pi/2$ and for all $u\in K_+$ we have
$||Fe_{(u,v)}||^2>1+\beta^2$.
\par Similarly,  for $\cos(v)\sin(v)<0$, let
$K_-:=\{u\in[0,\pi):\cos(2u+c)< \dfrac{\beta}{4\alpha}\}$. Then
for all $u\in K_-$, we have $||Fe_{(u,v)}||^2>1+\beta^2$ and
$|K_-|>\pi/2$.
\par Hence, given any $v\in[0,\pi)$, there exists an interval
 $K_v\subset[0,\pi)$, i.e. an interval in $\R\backslash\pi\Z$, such that $|K_v|>\pi/2$ and
 $||Fe_{(u,v)}||^2>1+\beta^2$, for all $u\in K_v$.
 Therefore, starting with an appropriately chosen vector $e_{(u,v)}\in\mathbf{P}(\C^2)$ such
 that $||Fe_{(u,v)}||^2>1+\beta^2$, applying $F$ will result in a vector
 $ce_{(u_1,v_1)}$ with $c>1$. Now we apply $E$  as many times as
 required to get a vector $ce_{(\tilde{u},v_1)}$ (or $-ce_{(\tilde{u},v_1)}$) such that
 $\tilde{u}\in K_{v_1}$. Iterating this process gives a sequence of
 vectors with unbounded norms. The case $\beta<0$ is treated similarly. Thus we have proved the  non-compactness of
  $\widetilde{H}_{\lambda,\mu}$, and consequently that of
  $\widetilde{G}_{\lambda,\mu}$.
  \end{proof}
  %%%%%%%%%%%%%%%%%%%%%%%%%%%%%%%%%%%%%%%%%%%%%%%%%%%%%%%%%%%%%%%%%%%%%%%%%%%%%%%%%%%%%%%%%%%%%%%

\par The next step is proving that $\widetilde{G}_{\lambda,\mu}$ is strongly
irreducible for all $\lm$ outside the set $M$.
\begin{Lemma}\label{t:d str.irr}
$\widetilde{G}_{\lambda,\mu}$ is strongly irreducible, for all
$\lm\in\mathbb{T}\backslash M$.
 \end{Lemma}
 \begin{proof}
  Since we already proved that $\widetilde{G}_{\lambda,\mu}$ is non-compact (Lemma~\ref{t:d
  noncomp}), it suffices to show that for all $v\in\mathbf{P}(\C^2),
\#\{gv:g\in \widetilde{G}_{\lambda,\mu}\}\geq 3$.
 \par We first note that $\rho^4=1$ implies that
 $|$tr$T(\theta,\theta)|\in\{0,2\}$, which in turn implies that $\lm\in\{-a,-b\}\cup M_a $.
  Therefore,
 the condition $\lm\notin \{-a,-b\}\cup M_a$ gives that
 $\rho^4\neq 1$. Hence, $\{I,E,E^2\}\subset
 \widetilde{G}_{\lambda,\mu}$ maps every $v\in\mathbf{P}(\C^2)$ to
 three different directions unless $v$ coincides with either $  \begin{pmatrix}
1\\
0\\
    \end{pmatrix}$ or $\begin{pmatrix}
0\\
1\\
    \end{pmatrix}$.
     At this point, we note that the diagonal elements of the matrix $F$, given in (\ref{F-matrix}), vanish simultaneously only if $\lm\in\{-a,-b\}$. Thus,
     iterations of the operator $F$, followed if necessary with iterations of
     $E$, maps each of the latter directions to at least three different
     elements of $\mathbf{P}(\C^2)$.
     This proves strong irreducibility of the group
 $\widetilde{H}_{\lambda,\mu}$, for all $\lm\notin\{-a,-b\}\cup M_a$,
  which gives the corresponding
 result for $\widetilde{G}_{\lambda,\mu}$.
      A similar argument, replacing the rules of $a$ and $b$,
      gives a similar assertion for
      $\lm\notin\{-a,-b\}\cup M_b$,
      thus finishing the proof.
\end{proof}
%%%%%%%%%%%%%%%%%%%%%%%%%%%%%%%%%%%%%%%%%%%%%%%%%%%%%%%%%%%%%%%%%%%%%%%%%%%%%%%%%%%%%%%%%%
\begin{proof}[\textbf{Proof of Theorem \ref{t:d lyp}}]
 The previous two lemmas combined with F\"urstenberg's Theorem, immediately
 give $\gm(\lambda)>0$ for all $\lm\in\mathbb{T}\backslash M$.
 Using Theorem~\ref{IP}, we deduce that
 $\Sigma_{ac}=\emptyset$.
\end{proof}
%%%%%%%%%%%%%%%%%%%%%%%%%%%%%%%%%%%%%%%%%%%%%%%%%%%%%%%%%%%%%%%%%%
\par  Even though the finiteness of the set $M$
is more than enough to prove the absence of absolutely continuous
spectrum for the unitary dimer model, $M$ is by no means optimal.
Determining whether or not a certain element of $M$ is, in fact, a
critical quasi-energy of $U_\og$ requires further analysis.
Nevertheless, the proof suggests that if the support of $\mu$
contains three or more points than generically $M=\emptyset$. Even
for the Bernoulli unitary dimer model where supp$\mu=\{a,b\}$ the
analysis is likely to fall into a number of different sub-cases.
However, for generic choices of $a,b$ we see that $M_a\cap
M_b=\emptyset$, thus $M=\{-a,-b\}$ and indeed the situation where
$\lm\in\{-a,-b\}$ is readily accessible. Guided by the proof of
Theorem 2.2(i) of~\cite{dg} for the self-adjoint dimer model we
prove that,
\begin{Proposition}
For a probability measure $\mu$ given by (\ref{bern.meas}), we
have the following
\begin{enumerate} [\upshape (i)]
\item If $|a-b|\in\sigma(S)$, then $\gm(-a)=\gm(-b)=0$.
\item If $|a-b|\in\rho(S)$, then both $\gm(-a)>0$ and $\gm(-b)>0$.
\end{enumerate}
\end{Proposition}
\begin{proof}
For $\lm=-a$, we have that $T(\theta,\theta)=-I$, and $\eta=b-a$.
Since
tr$\,T(\eta,\eta)=\dfrac{2r^2}{t^2}-\dfrac{2\cos(\eta)}{t^2}$,
then $|$tr$\,T(\eta,\eta)|\leq2$ when $b-a\in\sigma(S)$ and
$|$tr$\,T(\eta,\eta)|>2$ if $b-a\in\rho(S)$.
 \par Now, let
$m_n:=\#\{k:1\leq k\leq n,\theta_{2k}^\og=b\}$, then $\PP$-almost
surely
\[\lim_{n\to\infty}{\dfrac{m_n}{n}}=q.\]
This along with the fact that
\[\lim_{m_n\to\infty}{||[T(\eta,\eta)]^{m_n}||^{1/m_n}}= \max_{1\leq i\leq2}{|r_i|},\]
where $r_i$ are the eigenvalues of $T(\eta,\eta)$, gives the
results of the proposition.
\end{proof}

%%%%%%%%%%%%%%%%%%%%%%%%%%%%%%%%%%%%%%%%%%%%%%%%%%%%%%%%%%%%%%%%%%%%%%%%%%%%%%%%%%%%%%%%%%%%%%%%%%%%%%%%%%%%%%%%%%%%%%%%%%%%%%%%%%%%%%%%%%%%%%%%%%%%%%%%%%%%%%%%%%
%%%%%%%%%%%%%%%%%%%%%%%%%%%%%%%%%%%%%%%%%%%%%%%%%%%%%%%%%%%%%%%%%%%%%%%%%%%%%%%%%%%%%%%%%%%%%%%%%%%%%%%%%%%%%%%%%%%%%%%%%%%%%%%%%%%%%%%%%%%%%%%%%%%%%%%%%
%%%%%%%%%%%%%%%%%%%%%%%%%%%%%%%%%%%%%%%%%%%%%%%%%%%%%%%%%%%%%%%%%%%%%%%%%%%%%%%%%%%%%%%%%%%%%%%%%%%%%%%%%%%%%%%%%%%%%%%%%%%%%%%%%%%%%%%%%%%%%%%%%%%%%%%%%%
\section{Continuity of the Lyapunov Exponent} \label{contlyap}

 In this section we prove that, away from the critical points, the Lyapunov exponent is a continuous
 function of the spectral parameter $\lm$. The proof of this fact
 is similar to the one given in~\cite{cl} for the self-adjoint case.
 \par First, for a compact interval $I$ of quasi-energies with positive Lyapunov exponents, we define the function
 \[
 \Phi(\lm,\bar{v})=\E(\ln{\dfrac{||T_\lm
 v||}{||v||}}), \quad \bar{v}\in\mathbf{P}(\C^2),\lm\in I.
 \]
Where
$T_\lm:=T(\theta_{2n}^\omega(\lambda),\theta_{2n+1}^\omega(\lambda))$
denotes the transfer matrix defined in (\ref{trensfer matrices})
with the dependance on $\omega$ being suppressed in order to
simplify the notation. The next lemma establishes a couple of
properties of $\Phi(\lm,\bar{v})$.
\begin{Lemma}\label{Phi.cont.}
(i) The mapping $\bar{v}\mapsto\Phi(\lm,\bar{v})$ is continuous on
$\mathbf{P}(\C^2)$.
\\(ii) There exists a constant $C$ such that,
\[
\sup_{\bar{v}\in\mathbf{P}(\C^2)}|\Phi(\lm,\bar{v})-\Phi(\lm_1,\bar{v})|\leq
C|e^{i\lm}-e^{i\lm_1}|, \quad \lm,\lm_1\in I.
\]
\end{Lemma}
\begin{proof}
(i) From (\ref{trensfer matrices}) one sees that the norm of
$T_\lm$ is uniformly bounded for all $\lm,\omega$. Consequently,
we also have a uniform bound on $\Phi(\lm,\bar{v})$. The assertion
is then obtained using the dominated convergence theorem.
\par (ii) First we note that for all $\bar{v}\in\mathbf{P}(\C^2)$,
\begin{align*}
|\Phi(\lm,\bar{v})-\Phi(\lm_1,\bar{v})|&\leq\E(|\ln\dfrac{||T_\lm
v||}{||T_{\lm_1}v||}|).
\end{align*}
Since $|\det T_\lm T^{-1}_{\lm_1}|=1$, it follows that
\begin{align}\label{Phi}
|\Phi(\lm,\bar{v})-\Phi(\lm_1,\bar{v})|&\leq\E(\ln{||T_\lm
T^{-1}_{\lm_1}}||).
\end{align}
On the other hand, one has
\[||T_\lm T^{-1}_{\lm_1}||\leq ||T^{-1}_{\lm_1}||||T_\lm-T^{-1}_{\lm_1}||+1.\]
 Since all norms on $GL(2,\C)$ are equivalent, there exists a constant $C_1$ such that
 \begin{align}\label{Phi2}
 ||T_\lm T^{-1}_{\lm_1}||\leq C_1 ||T_\lm-T^{-1}_{\lm_1}||_F+1,
 \end{align}
 where $||A||^2_F\equiv
  \sum_{i,j=1}^2|a_{ij}|^2$ denotes the Frobenius norm of the matrix
 $A=\{a_{ij}\}_{i,j=1}^2$. From (\ref{trensfer matrices}), it is
 easy to see that $||T_\lm-T^{-1}_{\lm_1}||_F\leq
 C_2|e^{i\lm}-e^{i\lm_1}|$. Combining this with (\ref{Phi2}) and
 (\ref{Phi}) gives the required result.
\end{proof}
Before proving the main result of this section, we recall a
general fact: If $\mu_\lm$ denotes the probability measure on
$GL(2,\C)$ induced by $T_\lm$ and $G_{\lambda,\mu}$ is non-compact
and strongly irreducible for all $\lm\in I$, then there exists a
unique distribution $\nu_\lm$ on $\mathbf{P}(\C^2)$ that is
invariant with respect to $\mu_\lm$. A proof of this fact can be
found in~\cite{bl}. Moreover, we have
\begin{Lemma}\label{weakcont}
For $\lm\in I$, the mapping $\lm\mapsto \nu_\lm$ is weakly
continuous.
\end{Lemma}
\begin{proof}
We start by showing that if the sequence $\{\lm_n\}_{n\in\Z}$
converges to $\lm$, then the corresponding measures $\mu_{\lm_n}$
 converge weakly to $\mu_\lm$. First recall that for all $\omega\in\Omega$,
 $||T_{\lm_n}(\omega)-T_{\lm}(\omega)||\leq d_n$, where
 $d_n=C|e^{i\lm_n}-e^{i\lm}|$ for some $C\in \R$.
  Now, let $B\subset GL(2,\C)$ such that the boundary of
 $B$
 has zero measure with respect to $\mu_\lm$, i.e. $\mu_\lm(\partial B)=0$. For such a set we
 have that
 \begin{align*}
|\mu_{\lm_n}(B)-\mu_\lm(B)|&=|\PP[T^{-1}_{\lm_n}(B)]-\PP[T^{-1}_{\lm}(B)]|
\\&\leq\PP[T^{-1}_{\lm_n}(B)\cap
(T^{-1}_{\lm}(B))^c]+\PP[T^{-1}_{\lm}(B)\cap
(T^{-1}_{\lm_n}(B))^c]
 \end{align*}
It is not difficult to see that
\begin{align*}
\PP[T^{-1}_{\lm_n}(B)\cap (T^{-1}_{\lm}(B))^c]&\leq
\PP[\{\omega\in \Omega:T_{\lm}(\omega)\in B^c,
dist(T_{\lm}(\omega),\partial B)\leq d_n\} ]
\\&\leq \mu_\lm[\{A:dist(A,\partial B)\leq d_n\}]
\end{align*}
 Taking the limit as $n\to\infty$ and using dominated
 convergence one sees that
 \[\lim_{n\to\infty}\PP[T^{-1}_{\lm_n}(B)\cap
 (T^{-1}_{\lm}(B))^c]=\mu_\lm(\partial B)=0.\]
 Using a similar argument one gets that $\PP[T^{-1}_{\lm}(B)\cap (T^{-1}_{\lm_n}(B))^c]\to
 0$ as $n\to\infty$. Therefore, we have
 \[\lim_{n\to\infty}{\mu_{\lm_n}(B)}=\mu_{\lm}(B).\]
 Since this is true for any set $B$ with $\mu_\lm(\partial B)=0$, weak
 convergence of $\mu_{\lm_n}$ to $\mu_\lm$ follows~\cite{w}. In order to get the
 weak convergence of $\nu_{\lm_n}$, we use the fact that the set of
 invariant measures on $\mathbf{P}(\C^2)$ is compact in the weak*
 topology~\cite{w}, thus every subsequence of $\{\nu_{\lm_n}\}$ has a weakly
 convergent subsequence and since the limit of each of those
 subsequences is invariant with respect to $\mu_\lm$ it equals
 $\nu_\lm$ by uniqueness of the latter. A short contradiction
 argument shows that $\nu_{\lm_n}$ has to converge weakly to
 $\nu_\lm$.
\end{proof}
 The Lyapunov exponent can be expressed in terms
of the mapping  $\Phi$ and the measure $\nu_\lm$ as
\begin{align}
\gm(\lm)=\int\Phi(\lm,\bar{v}) d\nu_\lm(\bar{v}).
\end{align}
 Now we are ready to prove that for any interval $I$ for which $G_{\lambda,\mu}$ is non-compact
and strongly irreducible for all $\lm\in I$, we have
\begin{Theorem}
 The Lyapunov exponent  $\gm(\lm)$ is a continuous function of $\lm\in I$.
\end{Theorem}
\begin{proof}
Let $\lm\in I$ and choose $\{\lm_n\}\subset I$  a sequence of
quasi-energies such that $\lm_n\to\lm$ as $n\to\infty$. It follows
that
\begin{align}
\lim_{n\to\infty}{\gm(\lm_n)}&=\lim_{n\to\infty}{\int\Phi(\lm_n,\bar{v})\notag
d\nu_{\lm_n}(\bar{v})}
\\&=\lim_{n\to\infty}{[\int\Phi(\lm,\bar{v})
d\nu_{\lm_n}(\bar{v})+\int(\Phi(\lm_n,\bar{v})-\Phi(\lm,\bar{v}))
d\nu_{\lm_n}(\bar{v})]}\notag
\\&=\int\Phi(\lm,\bar{v})
d\nu_{\lm}(\bar{v}),\notag
\end{align}
 with the last equality following from Lemma~\ref{weakcont} along with part
 (ii) of Lemma~\ref{Phi.cont.}.
\end{proof}

\vskip .3cm

\noindent {\bf Acknowledgements:}  Partial financial support for
this project was provided through NSF grant DMS-0245210. We would
like to thank Alain Joye for useful discussions and Dirk
Hundertmark for providing the simple proof of Lemma~\ref{expect}.
E.~H.\ also acknowledges support through a Junior Research
Fellowship at the Erwin Schr\"odinger Institute in Vienna, where
part of this work was done.

%%%%%%%%%%%%%%%%%%%%%%%%%%%%%%%%%%%%%%%%%%%%%%%%%%%%%%%%%%%%%%%%%%%%%%%%%%%%%%%%

%%%%%%%%%%%%%%%%%%%%%%%%%%%%%%%%%%%%%%%%%%%%%%%%%%%%%%%%%%%%%%

%%%%%%%%%%%%%%%%%%%%%%%%%%%%%%

\end{document}